\documentclass[11pt,a4paper]{JHEP3}
\pdfoutput=1
\usepackage{graphicx}
\usepackage{bm}
\usepackage{bbm}
\usepackage{mathrsfs}
\usepackage{amsmath}
\usepackage{amssymb}
\usepackage{amstext}
\usepackage{yfonts}
\usepackage{epsfig,latexsym}
\usepackage{float}
\title{Physical Response Functions of Strongly Coupled Massive Quantum Liquids}

\author{
Bum-Hoon Lee $^{1,2}$, Xiaojian Bai $^{1,2}$, Matthias C. Wapler $^{1,3}$\\
{}\\
{\it $^1$ Center for Quantum Spacetime, Sogang University, Seoul, Korea}\\
{\it $^2$ Department of Physics, Sogang University, Seoul, Korea}\\
{\it $^3$ Since 05/2011: Department of Microsystems Engineering (IMTEK), University of Freiburg, Germany}\\
\\E-mail: \email{bhl@sogang.ac.kr}, \email{baixj@sogang.ac.kr
}, \email{matthias.wapler@imtek.uni-freiburg.de}}
\preprint{}
\newcommand{\tlrho}{{\tilde{\rho}}}
\newcommand{\tlc}{{\tilde{c}}}
\newcommand{\tlm}{{\tilde{m}}}

\newcommand{\tlt}{{\tilde{t}}} 
\newcommand{\tx}{{\tilde{x}}}
\newcommand{\ty}{{\tilde{y}}}
\newcommand{\tom}{{\tilde{\omega}}}
\newcommand{\bom}{{\bar{\omega}}}

\newcommand{\barm}{{\bar{m}}}

\newcommand{\Res}{\mathrm{Res}}
\newcommand{\order}{{\mathcal{O}}}

\renewcommand{\Re}{\mathrm{Re}}
\renewcommand{\Im}{\mathrm{Im}}
\newcommand{\labell}[1]{\label{#1}}
\newcommand{\reef}[1]{(\ref{#1})}
\date{\today}

\abstract{We study physical properties of strongly coupled massive quantum liquids from their spectral functions using the AdS/CFT correspondence. 

The generic model that we consider is dense, heavy fundamental matter coupled to $SU(N_c)$ super Yang-Mills theory at finite temperature above the deconfinement phase transition but below the scale set by the baryon number density. In this setup, we study the current-current correlators of the baryon number density using new techniques that 
employ a scaling behavior in the dual geometry.

Our results, the AC conductivity, the quasi-particle spectrum and the Drude-limit parameters like the relaxation time are simple temperature-independent expressions that depend only on the mass-squared to density ratio and display a crossover between a baryon- and meson-dominated regime. We concentrated on the (2+1)-dimensional defect case, but in principle our results can also be generalized straightforwardly to other cases.}

\begin{document}

\section{Introduction}
In the last decades, the AdS/CFT correspondence \cite{adscft_maldacena, adscft_gubser, adscft_witten} has become a powerful tool to study various properties of the strong coupling limit of conformal field theories, with applications to QCD and also more recently to some aspects of condensed matter physics. In principle, one has to distinguish between top-down setups that are constructed within string theory and imply consistency and bottom-up setups in which the gravitational duals are constructed from a phenomenological point of view. In this paper, we use the former approach as we would like to explore what happens to a particular consistent theory.

A common, and in particular top-down, approach how to introduce fundamental matter in AdS/CFT is the probe brane approach, where one considers a small number of ``probe branes'' in an $AdS$ black hole background which are dual to the fundamental matter coupled to an adjoint gauge theory. For example one considers the well-known $AdS_5 \times S^5$ (black hole) solution (above the deconfinement phase transition) that is a solution to a stack $N_c \gg 1$ D3-branes in the decoupling limit and is dual to a (thermal) $SU(N_c)$ $\mathcal{N} = 4$ supersymmetric Yang-Mills theory \cite{adscft_gubser}. Then one inserts $N_f \ll N_c$ intersecting Dp branes, giving $N_f$ families of (charged) fields in the fundamental representation of the $SU(N_c)$, living along the directions of the intersection \cite{winters,robfirst} - such as the above-mentioned D3-Dp intersections.

In condensed matter applications, there has been particular interest in $2+1$ dimensional systems that can be typically constructed using M2 branes \cite{pavel}, or as a defect in a $3+1$ dimensional background using D3-D5 \cite{conpaper, fancycon} and D3-D7 \cite{conpaper,fancycon,rey} intersections - and also as bottom-up setups in various contexts such as superconductivity. As our world is $3+1$ dimensional, the defect setup may be more realistic, even though there are some problems with the consistency of the D3-D7 setup \cite{conpaper, fancycon}. In recent years, there has also arisen significant interest in quantum-liquid-like aspects that arise when the temperature is small compared to the density, for example in a $3+1$ dimensional D3-D7 setup \cite{zerosound,funnel}, a $1+1$ dimensional D3-D3 configuration \cite{luttinger,funnel} or in a $2+1$ dimensional D3-D5 setup \cite{funnel}. 
In these systems several physical aspects have been studied, such as the zero-sound mode \cite{zerosound,starinets}, a fermionic instability \cite{bergman2011}, the quantum Hall effect \cite{jokela} and the response functions of fermionic matter at finite magnetic fields \cite{aoykhman}. This low-temperature limit, in particular the construction of \cite{funnel} will be in the focus of this paper

One interesting property of these configurations is a scaling behavior that occurs at small temperatures in setups that are dual to a finite density of fundamental matter with finite mass. Then one finds that solutions at different temperatures but fixed mass-squared to density ratios are equivalent through a simple scaling and display thermodynamic features reminiscent of quantum liquids \cite{funnel}. In this paper, we will exploit this scaling to further explore the properties of this state of matter through their two-point functions. This limit in parameter space corresponds to large quark number density with finite temperature or finite quark number density with small temperature but still above the deconfinement temperature, and any choice of quark mass in each case.

The paper is organized as follows: In section \ref{setup}, we review the setup and the formalism developed in \cite{funnel}. 
Then we will develop a method to solve the appropriate equations of motion in a temperature-independent parametrization in section \ref{solution} and obtain the physical results in section \ref{results}. There we first give an overview of the AC conductivity in section \ref{accon}, then obtain the small-frequency (Drude) limit in section \ref{drudesec} and the spectrum of quasiparticle excitations in section \ref{quasiparticle}. Finally we conclude in section \ref{conclude}.

\section{Setup}\label{setup}
The supergravity background of a planar black hole in $ AdS_{5} $ is 
\begin{eqnarray}\label{D3geom}
	ds^2 = \dfrac{r^2}{L^2}\left(-\left(1-\dfrac{r_0^4}{r^4}\right)\,dt^2 + d\vec{x}_3^{\,2}\right)+ \dfrac{L^2}{r^2}\left(\dfrac{dr^2}{1-r^4_0/r^4}+r^2 \,d\Omega_5^2\right),\quad C_{txyz}^{(4)} = -\dfrac{r^4}{L^4}.
\end{eqnarray}
This corresponds to the decoupling limit of $N_c$ black D3-branes
dual to ${\cal N}=4$ $SU(N_c)$ super-Yang-Mills theory at finite temperature $T$, living along the flat directions of the AdS \cite{adscft_witten}.
The  temperature $T$ is given by the Hawking temperature $T = \frac{r_0}{\pi L^2}$ and the Yang-Mills coupling by $g_{\text{YM}}^2 = 4 \pi g_s$. Since the curvature $L$ is given in terms of the string coupling $g_s$ and string length  $l_s$ as $L^4 = 4\pi\, g_s N_c \, l_s^4$, the 't Hooft coupling $\lambda=g_{\text{YM}}^2 N_c$ can be written as $\lambda=\frac{L^4}{l_s^4}$. Hence the ``supergravity limit'' $L \gg l_s$ in which the type IIB supergravity action and the solution \reef{D3geom} are valid  corresponds to the strong coupling $\lambda \gg 1$.

In practice, it is convenient to define dimensionless coordinates
\begin{eqnarray}
	u:=\dfrac{r_0}{r},&\quad \tilde{t}:=\dfrac{r_0t}{L^2},&\quad  \vec{\tx}:=\vec{x}\dfrac{r_0}{L^2}\,,
\end{eqnarray}
in which the metric $g$ becomes
\begin{equation}\label{branemetric}
ds^2 \, = \, \dfrac{L^2}{u^2} \left(-(1-u^4)d\tlt^2 + d\vec{\tx}_3^2 + \frac{du^2}{1-u^4} + u^2 d\Omega_5^2 \right) \ .
\end{equation}

In this setup all the fields transform in the adjoint representation of the $SU(N_c)$. In QCD or condensed matter physics, however, one would like to consider also matter that is charged under this symmetry, i.e. that transforms in the fundamental representation. Hence, the use of this solution for such applications is very much limited.
To introduce the fundamental matter one may create an intersection of ``probe'' Dp branes with the D3 branes, such that at the string theory side there are fields at the massless level of field theory at the intersection. From the point of view of the probe branes, they correspond to endpoints of D3-Dp strings, and in the field theory dual they correspond to fundamental fields of a (defect) field theory.

Here, we use the well-known D3-D5 defect setup (see e.g. \cite{lisa,hirosi,johannadef}):
\begin{equation}
\begin{array}{rccccc|c|cccccl}
  & & 0 & 1 & 2 & 3 & 4& 5 & 6 & 7 & 8 & 9 &\\
  & & t & x & y & z & r&   &   &   &   &  \theta &\\
\mathrm{background\,:}& D3 & \times & \times & \times & \times & & &  & & & & \\
\mathrm{probe\,:}& D5 & \times & \times & \times &  & \times  & \times & \times & &  & &  \ \ \ .
\end{array} 
\labell{array}
\end{equation}
The dual field theory is now the SYM gauge theory coupled to $N_f$ fundamental
hypermultiplets, which are confined to a (2+1)-dimensional defect.
This construction is still supersymmetric, but the supersymmetry has been reduced from ${\cal N}=4$ to
${\cal N}=2$ by the introduction of the defect. In the limit $N_f
\ll N_c$, the D5-branes may be treated as probes in the
supergravity background, i.e. we may ignore their gravitational
back-reaction.

We assume translational invariance along the flat directions and rotational invariance on the sphere. Hence, the pullback on D5-brane gives us one scalar field corresponding to the position in the $z$ direction, which was extensively studied in \cite{conpaper,fancycon}, and another scalar which describes the size of the compact sphere and corresponds to turning on the mass of the fundamental matter, studied in \cite{fancytherm,fancycon} and more extensively in the similar D3-D7 system in \cite{long,johanna,ingo,robdens,robchem}. Parametrizing the $S^5$ as 
$d\Omega_5^2 = d\theta^2 + \sin^2 \theta\, d\Omega_2^2 + \cos^2 \theta\, d\Omega_2^2$ and putting  the branes on the first $S^2$ of the $S^5$,
the induced metric $P[g]$ on the probe branes is given by
\begin{align}
	ds^2 &= \dfrac{L^2}{u^2}\left(-(1-u^4)d\tlt^2 +d\vec{\tx}_2^2 +\left(\dfrac{1}{1-u^4}+ \dfrac{u^2(\partial_u \Psi)^2}{1-\Psi^2}\right)du^2\right)\,\\
	 &+ L^2(1-\Psi^2)\,d\Omega_2^2\,, \notag
\end{align}
where we defined the scalar as $\Psi(u)=\sin(\theta(u))$.

We would like to turn on only the overall $U(1)$ factor of the world-volume gauge field, so the the probe branes are governed by the DBI action
\begin{eqnarray}\label{action}
	S(G) = - T_5 N_f \int_{D5}\sqrt{-\det G}\,, 
\end{eqnarray}
where $	G = P[g]+ 2\,\pi l_{s}^{2}F$.
This $U(1)$ gauge field is dual to the $U(1)$ current operator of the $U(N_f)$ that gives rise to the ``flavor symmetry''. In particular, the flux of the electric field
\begin{equation}
F \ = \ \partial_u A_{t}(u) \, du \wedge dt \ ,
\end{equation}
corresponds in the field theory side to the baryon or quark number density (see e.g. \cite{bigrev}), where we explicitly keep the quark number density $\rho_{\text{quark}} = N_c \rho_{\text{baryon}}$
\begin{align}\label{rhodic}
	\rho = \langle J^{t}\rangle = -\dfrac{\delta S}{\delta A_{t}^{bdy}}=4\pi N_f\dfrac{L^2T_5}{r_0}\lim_{u\rightarrow 0}\partial_u A_{t}(u)\,.
\end{align}

From the action \reef{action}, it is straightforward to obtain the solution for the gauge field
\begin{align}
	& &\partial_{u}A_{t}(u) = \tlrho \sqrt{\lambda} T  \dfrac{\sqrt{1-\Psi^2 + u^2(1-u^4)(\partial_{u}\Psi)^2}}{\sqrt{(1-\Psi^2)(\tilde{\rho}^2 u^4 + (1-\Psi^2)^2)}}\,,	
\end{align}
where $\tilde{\rho} =  \dfrac{2\rho}{N_c N_f T^2}$.
The equation of motion for $\Psi(u)$ is
\begin{align}\label{psieom}
	 &\partial_{u}\left(\partial_{u}\Psi\dfrac{1-u^4}{u^2}\sqrt{\dfrac{\tilde{\rho}^2  u^4+(1-\Psi^2)^2}{(1-\Psi^2)(1-\Psi^2+u^2(1-u^4)(\partial_{u}\Psi)^2)}}\right)\\ \notag
	&+ \frac{\Psi  \left(2 \left(1-\Psi^2\right)^3- u^2 (\partial_{u}\Psi)^2 \left(1-u^4\right) \left(u^4\tilde{\rho }^2-\left(1-\Psi^2\right)^2\right)\right)}{u^4 \sqrt{\left(1-\Psi^2\right)^3 \left(1-\Psi^2+(\partial_{u}\Psi)^2 \left(1-u^4\right) u^2\right) \left(u^4 \tilde{\rho }^2+\left(1-\Psi^2\right)^2\right)}} =0\,.
\end{align}

On the horizon $u=1$, this equation reduces to
\begin{equation}\label{psibdy}
	\lim_{u \rightarrow 1^{-}}\partial_u \Psi = \dfrac{1}{2} \dfrac{\Psi_0(1-\Psi_0^2)^2}{\tlrho^2+(1-\Psi_0^2)}, \quad \text{where}\quad \Psi_0 = \lim_{u \rightarrow 1^{-}} \Psi.
\end{equation}
The asymptotic solution at $u\rightarrow 0$ is 
\begin{align}\label{psiasym}
\Psi(u) = \tilde{m}\, u + \tilde{c} \, u^2 +\cdots\,,
\end{align}
where $\tlm$ and $\tilde{c}$ are dimensionless parameters that are determined by the value of $\Psi_0$ on the horizon.  Following arguments of the T-dual $(3+1)$ dimensional D3-D7 setup \cite{long,johanna,ingo,robdens,robchem}, the quark mass $M_q$ and dual condensate $C$ are given by 
\begin{equation}
M_q \, = \, \frac{\sqrt{\lambda}}{2}   T \tlm\ \ \  \mathrm{and}  \ \ \  C\, = \,   T^2 N_f N_c\tlc\ .
\end{equation}
One can understand this identification of the mass from the separation between the D3 and D5 branes in flat space, such that $M_q$ is the mass of a stretched D3-D5 string and the condensate is just the thermodynamic dual of the mass.

At vanishing density, there is a critical temperature-mass ratio below which the probe branes do not extend down to the horizon \cite{johannaphase,long}. At finite densities, unless one turns on the scalar in the $z$ direction considered in \cite{fancycon,fancytherm}, the branes always extend down to the horizon even though a phase transition may still be observed at small densities \cite{fancytherm}. In the limit considered in this paper, however, this phase transition is of no concern.

To obtain the low-temperature scaling solutions, we make a transformation $u=\xi/\sqrt{\tlrho}$ for \reef{psieom} and expand it in the large-$\tlrho$ (i.e. $\rho \ll T^2$) limit. The leading order gives
\begin{align}\label{psieomxi}
&\partial _{\xi }\left(\frac{\partial _{\xi }\Psi  }{\xi ^2}\sqrt{\frac{\xi ^4+\left(1-\Psi ^2\right)^2}{\left(1-\Psi ^2\right) \left(1-\Psi ^2+\left(\partial _{\xi }\Psi \right){}^2 \xi ^2\right)}}\right)\notag\\
&+ \frac{\Psi  \left(2 \left(1-\Psi ^2\right)^3-\left(\partial _{\xi }\Psi \right){}^2 \xi ^2 \left(\xi ^4-\left(1-\Psi ^2\right)^2\right)\right) }{\xi ^4 \left(1-\Psi ^2\right) \sqrt{\left(1-\Psi ^2\right) \left(1-\Psi ^2+\left(\partial _{\xi }\Psi \right){}^2 \xi ^2\right) \left(\xi ^4+\left(1-\Psi ^2\right)^2\right)}}=0\,,
\end{align} 
which doesn't depend on $\tlrho$ explicitly and there remains only implicit dependence from the scaling of $\xi$.
In the large $\xi$ limit, we have a series expansion solution
\begin{align}\label{psibdyxi}
	\Psi(\xi)=\Psi_0-\frac{\Psi_0\left(1-\Psi_0^2\right)^2}{10\xi ^4} +\cdots\, ,
\end{align}
which serves as the boundary condition to solve \reef{psieomxi} numerically. 
In the small $\xi$ limit, we obtain 
\begin{align}
\Psi(\xi) = \bar{m}\, \xi +\bar{c}\, \xi ^2 + \cdots\, .
\end{align}
The parameters $\bar{m}$ and $\bar{c}$ are now related to the quark mass, condensate  and density  by
\begin{align}
	\bar{m} =\dfrac{\tilde{m}}{\sqrt{\tlrho}} = M_q \sqrt{\dfrac{2 N_c N_f}{\rho\lambda}}\quad \text{and} \quad
	\bar{c} =\dfrac{\tilde{c}}{\tlrho} = \dfrac{2C}{\rho}\,.
\end{align}

In general, equation \reef{psieomxi} has no analytic solution and \reef{psibdyxi}  implies that in practice one has to start integrating the equation from the large-$\xi$ region -- that replaced the horizon boundary condition -- to obtain the mass and the related condensate at the asymptotic boundary, rather than setting either of them in the beginning.

\section{Solving the Fluctuation Field}\label{solution}
To obtain the conductivity though linear response theory and also the spectrum of excitations, we need to know the two-point functions of current operators. In AdS/CFT, they are obtained from the variation of the on-shell action --
since the baryon number current $J_i$ is dual to the gauge field of the $U(1)$ subgroup of the $U(N_f)$, the variation with respect to $A_i$:
\begin{eqnarray}\labell{correlact}
C_{i j} & =& \frac{\delta^2 S}{\delta A^\star_{i, 0}\delta A_{j, 0}}\\  \mathrm{where} \ \
C_{ij}(x-y) & = & -i\,\theta(x^0-y^0)\,\langle\, [J_i(x),J_j(y)]\, \rangle
\end{eqnarray}
and $A_{i,0}$ is the boundary value of the gauge field at the asymptotic boundary $u=0$.
Hence we perform a pertubation $ F \rightarrow F + f $ of the gauge field and expand the action up to quadratic order,
\begin{align}
S & = - T_5 N_f \int_{D5}\sqrt{-\det(G + f)}\nonumber\\
   & \approx S(G) + S^{(2)}(G,f)\, ,
\end{align}
where
\begin{align}\label{pertact}
S^{(2)}(G,f)= \dfrac{1}{4}T_5 N_f \int_{D5} \sqrt{-\det G}\, (G^{\alpha \mu}G^{\nu \beta }f_{ \beta \alpha}f_{\mu \nu}-\dfrac{1}{2} G^{\mu \nu} f_{\mu \nu}  G^{\alpha \beta}f_{\alpha \beta})\,,
\end{align}
where $G = P[g]+ 2\,\pi l_{s}^{2}F$.
The variation of eq. (\ref{correlact}) then becomes
\begin{equation}\labell{condmode}
C_{i j} = - \varepsilon_0 \left. \frac{A_j'(u)}{A_{i}(u)}\right|_{u\rightarrow 0} \ , \ \ \varepsilon_0 = 2T\frac{N_f N_c}{\sqrt{\lambda}}
\end{equation}
and we see that all we need to do is to solve the equations of motion for the gauge field obtained from eq. (\ref{pertact}) in the background of sec. \ref{setup} which is all the usual well-known method.

To obtain temperature-independent expression in the spirit of \cite{funnel}, we do this however in a smart manner that represents the temperature independence. To do so, we obtain a solution at large radius $u \ll 1$ or $\xi \ll \sqrt{\tlrho}$ and fix it in an overlap in the region $1/\sqrt{\tlrho} \ll u \ll 1$ or $\sqrt{\tlrho} \gg \xi \gg 1$, with an analytic solution in terms of the original radial coordinate $u$ that is valid at relatively small radius $1/u \ll \sqrt{\tlrho}$ or $\xi \gg 1$, and reflects the horizon boundary condition.

We fix the gauge to $ A_{u}=0 $, and we work in the case of vanishing momentum where $ \partial_{\tx} A_{\mu}=0$ and $\partial_{\ty}A_{\mu}=0 $ and the x and y component equations are the same. As discussed in \cite{conpaper,fancycon} in the context of the $SL(2,\mathbb{Z})$ duality, the $t$-component is also related to the spatial components, such that we only need to obtain the solution for $A_{\tx}$. After Fourier transformation, the x-component equation of motion is 
\begin{align}\label{axueom}
&\partial_{u} \left( (1-u^{4})\sqrt{\dfrac{((1-\Psi^2)^{2}+u^4\tilde{\rho}^2)(1-\Psi^2)}{(1-\Psi^2+(\partial_{u} \Psi)^{2}u^{2}(1-u^{4}))}} \partial_{u} A_{\tx} (\tilde{\omega},u)\right) \notag \\ 
& + \dfrac{1}{(1-u^{4})} \sqrt{\dfrac{(1-\Psi^{2}+(\partial_{u} \Psi)^{2}(1-u^{4})) ((1-\Psi^{2})^{2}+u^{4}\tilde{\rho}^{2})}{(1-\Psi^{2})}} \tilde{\omega}^{2} A_{\tx} (\tilde{\omega},u)=0\,,
\end{align}
where $\tom = \frac{\omega}{\pi T}$.
In the near horizon region, \reef{axueom} is reduced to 
\begin{align}\label{axubdy}
	\epsilon \partial_{\epsilon}(\epsilon \partial_{\epsilon} A_{\tilde{x}}(\tilde{\omega},u))+\dfrac{ \tilde{\omega}^2}{16}A_{\tx} (\tilde{\omega},u)=0
\end{align}
where $\epsilon=1-u$. The solution is 
\begin{align}\label{axubdysol}
	A_{\tx} (\tom,u)
	 =  C_{+}(1-u)^{+ i \tilde{\omega}/4} + C_{-}(1-u)^{- i \tilde{\omega}/4}\,,
\end{align}
where $ C_{\pm} $ are integration constants. The in-falling boundary condition only picks up the second term. We set $C_-$ to be 1 for convenience in the later discussion, which will not affect the conductivity and other physical results.

First let us obtain the solution for $1/u \ll \sqrt{\tlrho}$. In the limit $\tilde{\rho} \gg u^{-2}$, \reef{axueom} reduces to 
\begin{align}\label{axurho}
 \partial_{u} \left( u^{2}(1-u^{4}) \partial_{u} A_{\tx} (\tilde{\omega},u) \right) + u^{2} \tilde{\omega}^{2} A_{\tx} (\tilde{\omega},u)+ \order (\tlrho^2 u)^{-2} =0\, ,
\end{align}
where we made use of the scaling in eq. (\ref{psibdyxi}).
To solve this equation, we take the Ansatz
\begin{align}
A_{\tx} (\tilde{\omega},u)=C \exp^{\int_{u^\star} ^{u} \zeta(u') du'}\,,
\end{align}
with some arbitrary number $u^\star$, the E.O.M. in this limit becomes
\begin{align}
u(-1+u^{4})^{2}\zeta(u)^{2}+ (2-8u^{4}+6u^{8})\zeta(u)+ u\tilde{\omega}^{2}+u(-1+u^{4})^{2}\zeta'(u)=0\,.
\end{align}
This equation can be solved order by order in the limit of large $\tom$ (i.e. $\omega \gg T$) and we obtain the leading terms
\begin{align}
\zeta(u) = \dfrac{i \tilde{\omega}}{1-u^4}-\dfrac{1}{u}+\cdots\, .
\end{align}
Hence the solution for $A_x( \tom,u)$ is
\begin{align}\label{axurhosol}
A_{\tx} (\tilde{\omega},u)  = a_0(\tom) \dfrac{1}{u}\exp( - i \tilde{\omega }\frac{1}{4}( \ln[1-u] -\ln[1+u] -2\arctan[u]))\,,
\end{align}
where $a_0(\tom) = \exp(-i \tilde{\omega }\frac{1}{4}\left( \ln[2]+\frac{\pi }{2}\right))$ which is obtained by matching \reef{axubdysol} and \reef{axurhosol} in the near horizon limit.

Next, we consider the E.O.M. in the outer region $\xi \ll \sqrt{\tlrho}$.
After transforming \reef{axueom} into $\xi$-coordinate, in the large $\tlrho$ expansion the leading order gives the simplified E.O.M.
\begin{align}\label{axxieom}
\alpha (\xi )\partial_{\xi}\left(\beta (\xi ) \partial_{\xi}A_{\tilde{x}}(\bom, \xi )\right)+\bar{\omega }^2 A_{\tilde{x}}(\bom, \xi) + \order (\xi^2/\tlrho)^2=0\,,
\end{align}
where
\begin{align}
	\alpha (\xi )&=\sqrt{\frac{1-\Psi^2}{\left(1-\Psi ^2+(\partial_{\xi}\Psi)^2 \xi ^2\right) \left(\xi ^4+\left(1-\Psi^2\right)^2\right)}}\,,\\
	\beta (\xi )&=\frac{\sqrt{\left(\xi ^4+\left(1-\Psi^2\right)^2\right) \left(1-\Psi^2\right)}}{\sqrt{1-\Psi^2+(\partial_{\xi}\Psi)^2 \xi ^2}}\,,
\end{align}
and  $\bom =\tom /\sqrt{\tlrho}$. Note that the explicit $\tlrho$ dependence has dropped out, which is the advantage of the $\xi$-coordinate.
In the large $\xi$ limit, this equation is further simplified to 
\begin{align}\label{axxibdy}
	\xi ^2\partial _{\xi }\left(\xi ^2\partial _{\xi }A_{\tx}(\bar{\omega },\xi)\right)+ \bar{\omega }^2\xi ^4A_{\tx}(\bar{\omega },\xi) = 0\,,
\end{align}
and the solution is 
\begin{align}\label{axxibdysol}
A_{\tx}(\bom,\xi)=C_1\frac{e^{-i \xi  \bar{\omega }}}{\xi }+C_2\frac{e^{+i \xi  \bar{\omega }}}{\xi }\,.
\end{align}
where $C_1=0 $ and $ C_2 = \sqrt{\tlrho}e^{-i \bom \sqrt{\tlrho}\frac{1}{4}\left( \ln[2]+\frac{\pi }{2}\right)}$ are fixed by matching this solution with \reef{axurhosol} in the $1/\sqrt{\tlrho} \ll u \ll 1$ region. We can take \reef{axxibdysol} as the boundary condition to solve \reef{axxieom} numerically, replacing what would be usually the boundary condition on the horizon.

\begin{figure}[H]
\centering
\includegraphics[width =0.95\textwidth]{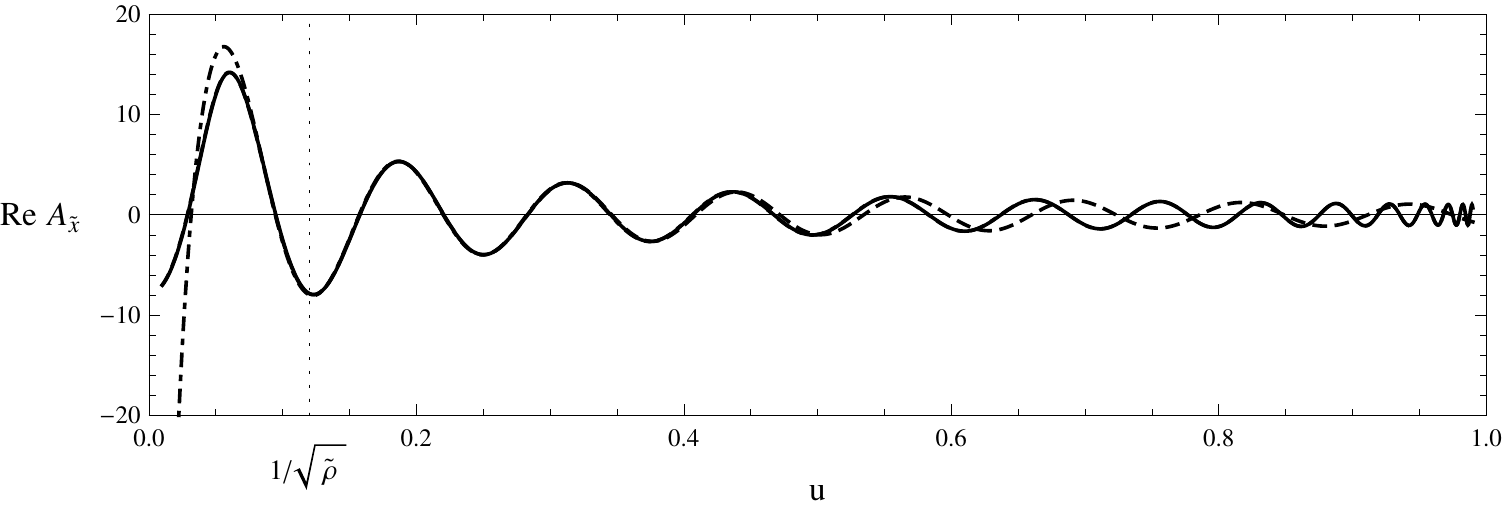} 
\caption{Comparison of the numerical solution for the full E.O.M. \protect \reef{axueom} (solid), the approximate solution \protect \reef{axurhosol} (dot-dashed) and the numerical solution of simplified E.O.M. \protect\reef{axxieom} (dashed) with $\tlrho = 70$, $\tom = 50$ (or $\bar{\omega} = 5.98$) and $\tlm = 34.1$ ($\Psi_0 = 0.9$, $\bar{m} = 4.08$). The dotted vertical line indicates the location of $1/\sqrt{\tlrho}$.  }\label{axufig}
\end{figure}
In figure \ref{axufig}, we see an example how the solution to the full equations,  eq. \reef{axueom}, the ``near horizon'' solution \reef{axurhosol} and the solution to \protect\reef{axxieom} agree in the regions $u \gg 1/\sqrt{\tlrho}$ and $u \ll 1$, respectively -- and in the overlapping region $1/\sqrt{\tlrho} \ll u \ll 1$. 
\section{Physical results}\label{results}
In this section, we give an overview of some interesting properties that we can extract from the correlator eq. (\ref{correlact}). For the transport properties, we obtain the conductivity from linear response theory, i.e. using the Kubo formula 
\begin{align}\label{kubo}
	\sigma_{ij} = \dfrac{i}{\omega} C_{ij}\, .
\end{align}

\subsection{AC Conductivity}\label{accon}
\begin{figure}[H]
\centering
\includegraphics[width = 0.60\textwidth]{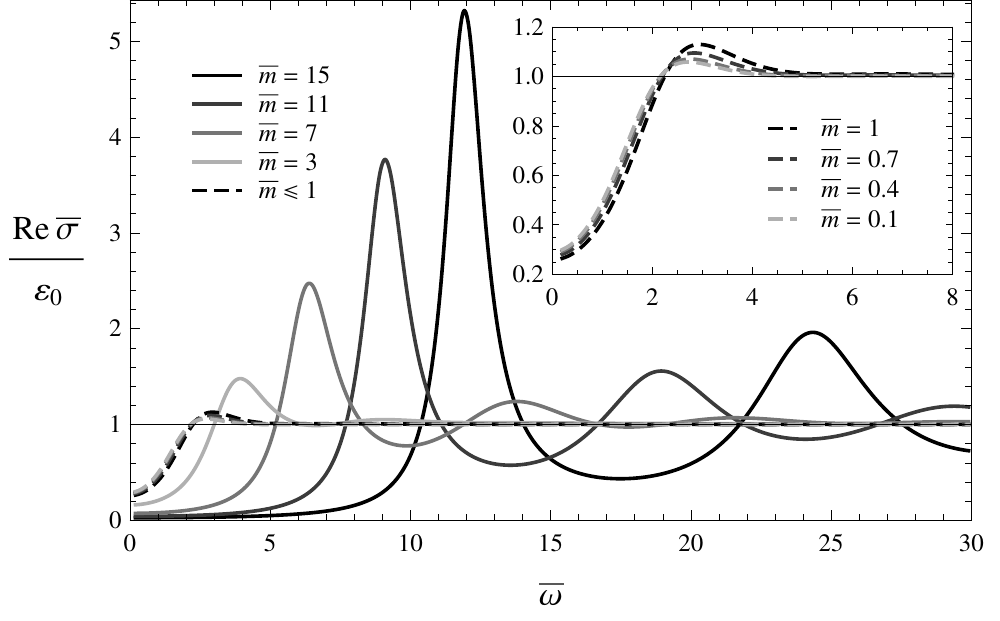}
\caption{The real part of AC conductivity with various values of $\barm$. }\label{ACfig}
\end{figure}

First let us look for an overview of the AC conductivity.
Defining for convenience $\bar{\sigma}_{ij} = \pi T \sigma_{ij}$, eqs. (\ref{kubo}) and (\ref{condmode}) give us
\begin{align}\label{conduc}
	\bar{\sigma}= - \varepsilon_0 \dfrac{i}{\bom}\dfrac{A_{\tx}'(\xi)}{A_{\tx}(\xi)}\Big |_{\xi\rightarrow 0}\,, \ \ \ \bom = \dfrac{\omega}{T\pi\sqrt{\tlrho}}.
\end{align}
We should note that the conductivity doesn't explicitly depend on $\tlrho$, since it appears only in the pre-factor of $A_{\tx}(\xi,\bom)$ and gets canceled out in the conductivity.
The AC conductivity for various $\bar{m}$ and $\bom$ is shown in the fig. \ref{ACfig}. 
We see at large frequencies the ``plasma'' resonances that were discussed in \cite{fancycon}, and at small frequencies the absence of the Drude-peak. The latter is due to the fact that we took the limit $\omega \gg T$ to obtain \reef{axurhosol}, and we will reconstruct it in the following section. We will also discuss the spectrum of ``plasmons'' that gives rise to the resonances in section \ref{quasiparticle}.

\begin{figure}[H]
\centering
\includegraphics[width = 0.49\textwidth]{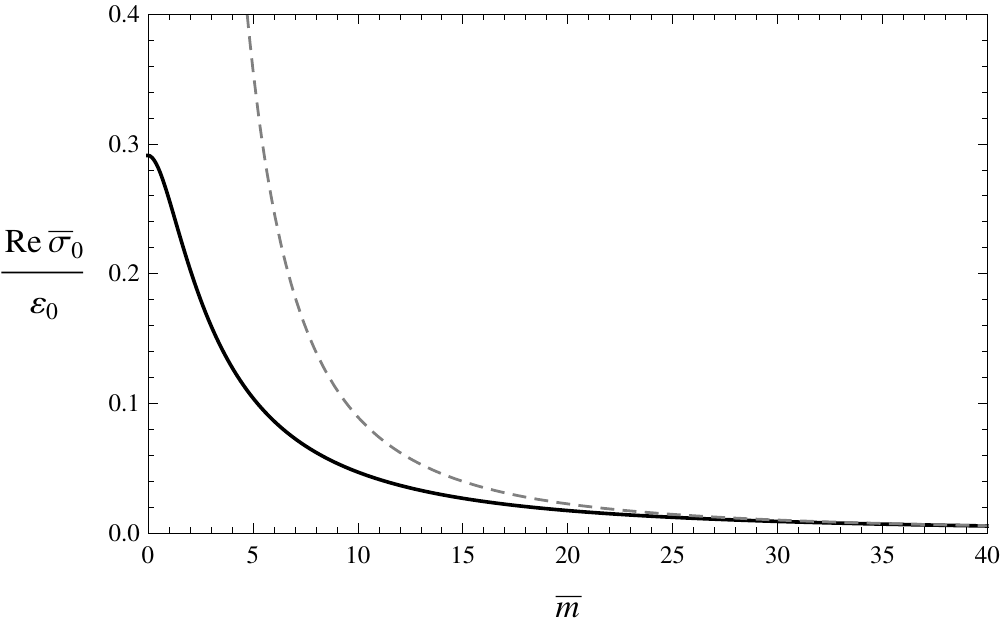} 
\hspace{0\textwidth}
\includegraphics[width = 0.49\textwidth]{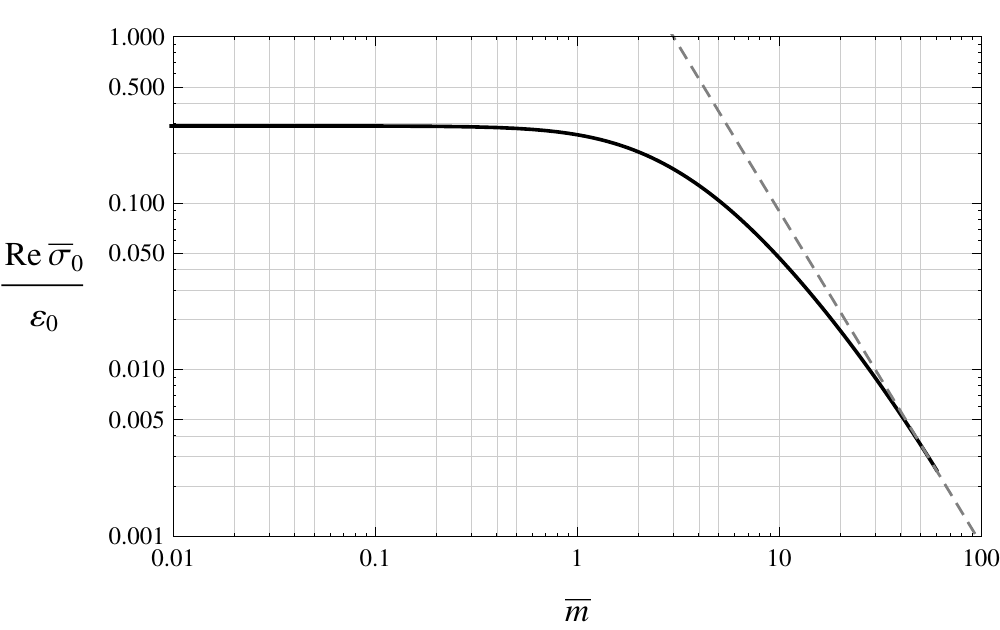} 
\caption{The conductivity $\bar{\sigma}_0$ as a function of $\bar{m}$. The dashed line indicates the asymptotic behavior $\mathrm{Re} (\bar{\sigma}) \approx 8.94 \varepsilon_0 / \bar{m}^2$. }\label{zerofreq}
\end{figure}

An interesting observation is that the temperature-scaled AC conductivity goes to some finite value $\bar{\sigma}_0$ (i.e. the conductivity scales as $\bar{\sigma}_0/(\pi T)$) as the frequency $\bom$ approaches 0 (but still above the Drude peak), which we plot in fig. \ref{zerofreq} as a function of $\barm$.
This constant, $\frac{\pi T}{\varepsilon_0}\Re(\sigma_0)$, approaches the constant 0.29 in the small $\barm$ region, and vanishes as $\mathrm{Re} (\bar{\sigma}_0) \approx 8.94 \varepsilon_0 / \bar{m}^2$  when $\barm$ is large. Physically, it reflects an accumulated contribution from the Drude peak and all the plasmon modes, and the scaling implies that they contribute less as the quark mass becomes much larger than the scale set by the density -- which is in line with the fact that the resonances become more stable, i.e. narrower with increasing mass as we will find out in section \ref{quasiparticle}. 
\subsection{Relaxation Time in the Drude limit}\label{drudesec}
The Drude model describes usually the DC and small-frequency conductivity of weakly coupled systems, such as electrons in simple metals. In \cite{fancycon} it was found, however, that the Drude conductivity given by (see e.g. \cite{snokebook})
\begin{align}\label{drude}
\sigma  =\frac{\sigma _{\text{DC}}}{1-i \omega \tau}\,
\end{align}
also applies to strongly coupled systems and describes the conductivity very well in a consistent manner, up to a ``conformal'' contribution that is subleading at large densities. 
While the relaxation time was given only numerically, the DC conductivity was found to be
\begin{align}\label{dccon}
\sigma _{\text{DC}}= \frac{\varepsilon_0}{\pi  T}\sqrt{\left(1-\Psi_0^2\right){}^2+\tilde{\rho }^2}\approx \frac{\varepsilon_0\tilde{\rho }}{\pi  T}\, .
\end{align}
For convenience, we define the relaxation time $\bar{\tau} = \tau (T \pi \sqrt{\tlrho})$ such that $\omega \tau = \bar{\omega}\bar{\tau}$.

To obtain a closed-form expression for $\bom \ll 1$, we look for an approximate solution to \reef{axxieom} by considering only the first term.
The solution for $A_{\tx}(\xi)$ is then
\begin{align}\label{axxibom}
	A_{\tx}(\xi) = \int_{+\infty}^{\xi}\dfrac{C_1}{\beta(\xi')}\,d\xi' + C_2\, . 
\end{align}

In the region $0\ll \xi \ll 1/\bom$ region, $\beta(\xi)\approx \xi^2$ such that matching \reef{axxibom} with \reef{axxibdysol} in this region gives us $C_2 = 0$. $C_1$ cancels out in  eq. (\ref{conduc}), such that we obtain  the conductivity 
\begin{align}\label{DCapprox}
	\dfrac{\bar{\sigma}}{\varepsilon_0} = -  \dfrac{i}{\bom} \left(\int_{+\infty}^{\xi}\beta^{-1}(\xi')\,d\xi'\right)^{-1}\Big |_{\xi\rightarrow 0}\, .
\end{align}
This is purely imaginary and has a pole at $\bom = 0$ with the residue 
\begin{align}\label{resiint}
	\Res_0(\dfrac{\bar{\sigma}}{\varepsilon_0}) = -   \left(\int_{+\infty}^{\xi}\beta^{-1}(\xi') \,d\xi'\right)^{-1} \Big |_{\xi\rightarrow 0}\,. 
\end{align}

%
%
%
%
%
%
%
%
%

%
Due to the Kramers-Kronig relations, the real part of conductivity becomes a $\delta$-function in $\bom$ with a pre-factor $\pi \Res_0(\dfrac{\bar{\sigma}}{\varepsilon_0})$ which we will compare to the Drude conductivity (\ref{drude}) by only considering the area under the peak, as our methods do not ``resolve'' its detailed frequency dependence.
To do so, we integrate the real part of the conductivity eq. (\ref{drude}) over the frequency
\begin{align}
\int_{-\infty }^{+\infty }\Re(\sigma)\,d\bom=\int_{-\infty }^{+\infty } \frac{\sigma _{\text{DC}}}{1+\bom ^2 \bar{\tau} ^2}\,d\bom = \frac{\pi\sigma _{\text{DC}}}{\bar{\tau }}\, 
\end{align}
and identify this with the integral over the frequency in the low-frequency regime of our low-temperature-limit result, i.e. with the weight of the delta function. This is justified by the fact that our scaling and approximation $\tom \gg \sqrt{\tlrho}$ has the effect of ``shrinking'' the Drude-region to zero size in $\bar{\omega}$. Using expression for the DC conductivity from \cite{fancycon} (\ref{dccon}) and the expression for the residue, (\ref{resiint}), we can obtain an expression for the relaxation time
%
%
\begin{align}
\bar{\tau} 
=  \frac{2\rho}{N_c N_f T^2}\int_{\xi}^{+\infty}\beta^{-1}(\xi') \,d\xi' \Big |_{\xi\rightarrow 0} \ .
\end{align} 
At $\bar{m} = 0$, this result can be solved analytically as $\beta \stackrel{\bar{m}=0}{=} \sqrt{1+\xi^4}$ such that
\begin{align}
\bar{\tau} = \frac{2\rho}{N_c N_f T^2} \int_{0}^{+\infty}\frac{d\xi}{\sqrt{1+\xi^4}}  =  \frac{\Gamma(1/4)^2}{2\sqrt{\pi}}\frac{\rho}{N_c N_f T^2}  
\end{align}
%
\begin{figure}[H]
\includegraphics[width = 0.49\textwidth]{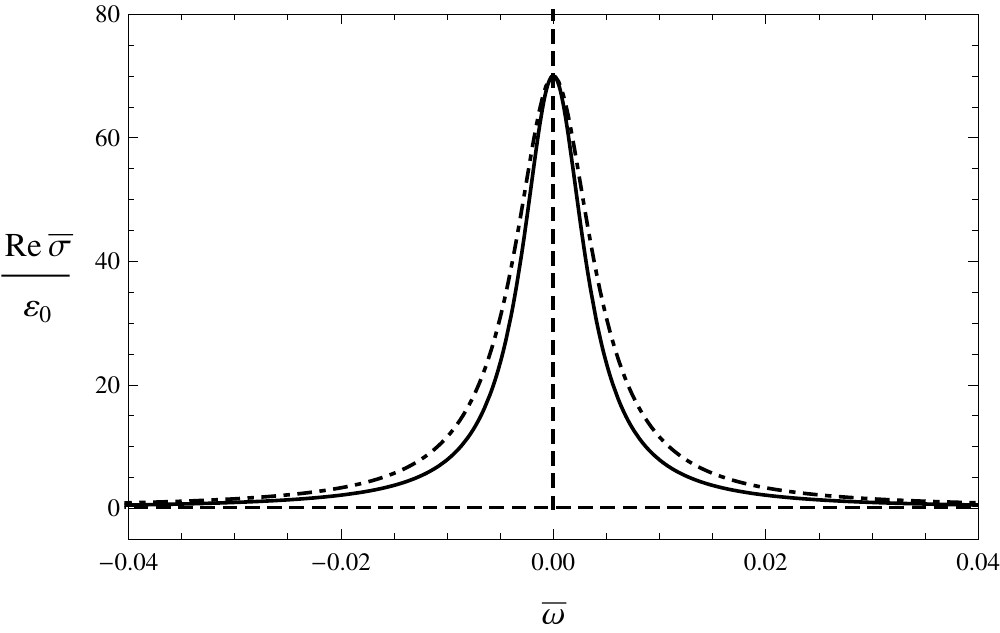} 
\hspace{0\textwidth} 
\includegraphics[width = 0.49\textwidth]{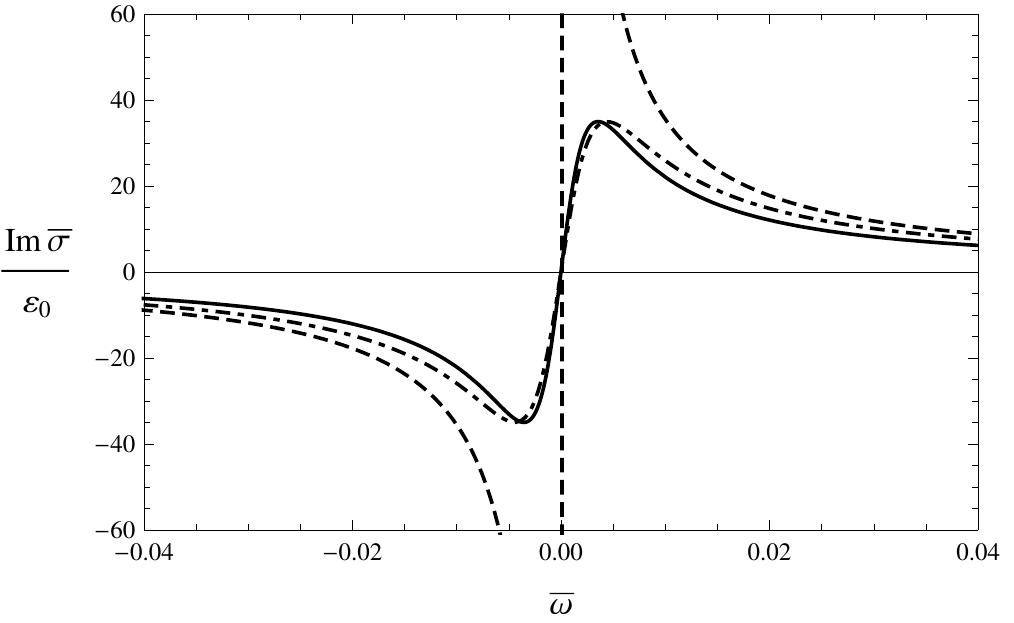} 
\caption{The comparison of the conductivities obtained from a) the Drude conductivity formula \protect \reef{drude} (solid) b) our approximate small-temperature solution \protect \reef{DCapprox} (dashed) and c) the solution of the full equation \protect \reef{axueom}, with $\tlrho = 70$\,, $\barm = 4.08$ and accordingly $\tau = 196.72$.
}\label{dccompare}
\end{figure}

In fig. \ref{dccompare}, we compare the Drude conductivity with our numerically-obtained relaxation time to the result obtained from the full equations of motion and the result from our low-temperature limit for an illustrative large value $\tlrho = 70$ and finite value $\bar{m} = 4.08$. We see reasonably close agreement. Already for $\tlrho = 100$, there would be no difference visible between the result for the full conductivity and the one from the Drude conductivity.

\begin{figure}[H]
\centering
\includegraphics[width = 0.5\textwidth]{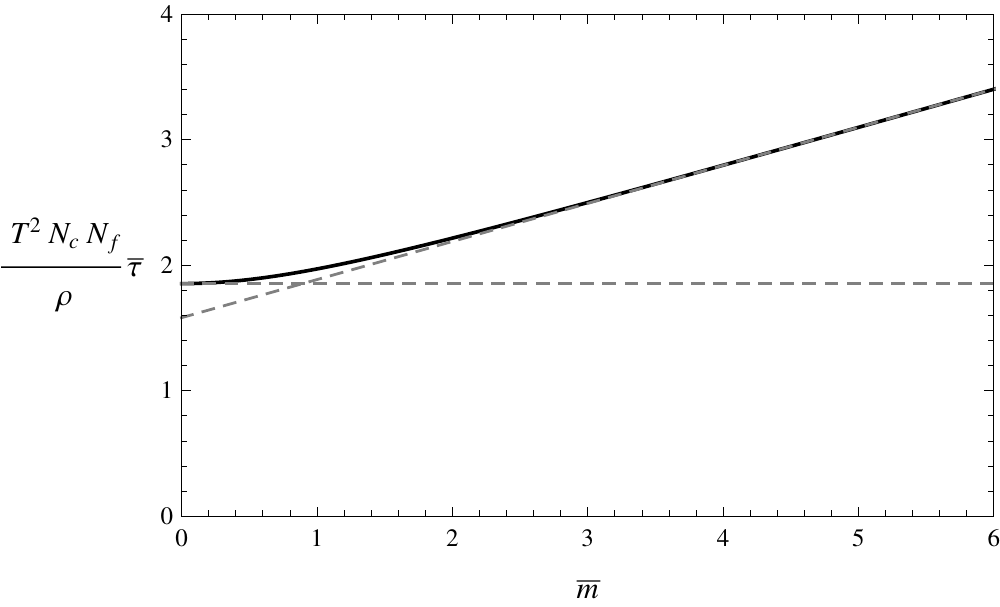}
\caption{The relaxation time as a function of quark mass, $\bar{m}= \frac{M_q}{\sqrt{\rho}}\sqrt{\frac{2N_c N_f}{\lambda}}$. In the small $\barm$ region, $\bar{\tau} = \tau \pi \sqrt{\frac{2 \rho}{ N_c N_f}} $ approaches a constant, $\frac{\Gamma(1/4)^2 \rho}{2\sqrt{\pi} N_c N_f T^2}$,  approximately 1.85, while it increases linearly with a slope of 0.27 when $\barm$ is large.}\label{drudeplot}
\end{figure}

In fig. \ref{drudeplot}, we show the relaxation time as a function of the quark mass. At small $\barm$, i.e. when the system is dominated by its density, 
we find  $\tau \propto\frac{1}{T^2} \sqrt{\frac{\rho}{ N_c N_f}}$. From a weak-coupling ``mean free path'' point of view, this is somewhat counter intuitive, but it can be understood as the system is strongly coupled and the conductivity is due to collective excitations that become more stable as the density is increased. At large quark mass, we have $\tau \propto \frac{M_q}{\sqrt{\lambda}T^2}$ which is independent of the density as we are in the limit were the shortest length scale is the Compton wavelength of the quark mass (or more precisely of the energy scale of mesons \cite{robfirst} $\frac{M_q}{\sqrt{\lambda}}$ that dominates the type of system at large quark mass \cite{long,robdens,robchem,fancycon}). While this impression implies that the relaxation time increases with increasing mass and decreasing temperature, it is not of the form of the weak-coupling geometric scattering. Also the factor of the meson mass might be misleading, as they cannot carry the baryon number current.

\subsection{Quasiparticle Spectrum}\label{quasiparticle}
The spectrum of quasiparticles corresponds to the locations of the poles of the correlator in the complex frequency plane, and hence to quasinormal modes in the gravity side. Thus, it can be obtained by tuning the complex frequency $\bom$ such that the fluctuation fields vanish at the asymptotic boundary \cite{alexandpaolo}. It is straightforward to search the exact locations of poles for large $\bar{m}$ by directly solving \reef{axxieom} for complex $\bom$ numerically, but the numerics start to break down when we approach to the small $\barm$ region, where we can only obtain few first order poles. Hence, to study for higher order excitations and small values of $\barm$ we developed a method based on analytic continuation.

\subsubsection{Analytic Continuation}\label{analcon}
Our general strategy is to find a closed-form expression that is exact for large $|\bom| \gg 1$. As this expression numerically still does not converge for sufficiently large imaginary parts of $\bom$, we then do a parametric fit for real frequencies and then perform an analytic continuation to obtain the location of the poles in the complex frequency plane.
First, we take the Ansatz
\begin{align}\label{aintansatz}
	A_{\tx}(\xi) = C e^{\int^{\xi}_{\Lambda}\zeta(\xi')\,d\xi'}\,,
\end{align}
to re-cast \reef{axxieom} into
\begin{align}\label{modeeom}
\zeta (\xi )^2+\frac{\beta '(\xi )}{\beta (\xi )}\zeta (\xi )+\zeta '(\xi )+\frac{\bar{\omega }^2}{\alpha (\xi ) \beta (\xi )}=0\,.
\end{align}
Then, we let $\zeta(\xi) = \zeta_0(\xi)+\zeta_1(\xi)$, where $\zeta_0(\xi)$ is the leading term in $\bom^{-1}$ and satisfies the algebraic equation
\begin{align}
	\zeta _0(\xi ){}^2+\frac{\beta '(\xi ) }{\beta (\xi )}\zeta _0(\xi ) + \frac{\bar{\omega }^2}{\alpha (\xi ) \beta (\xi )}=0\,,
\end{align}
with the solutions
\begin{align}
\zeta _0(\xi )=-\frac{\beta '(\xi )}{2 \beta (\xi )}\pm i \sqrt{\frac{\bar{\omega }^2}{\alpha (\xi ) \beta (\xi )}-\frac{\beta '(\xi )^2}{4 \beta (\xi )^2}}\,.
\end{align}
The subleading term $\zeta_1(\xi)$ should thus satisfy 
\begin{align}\label{zetaeom}
\zeta_1 '(\xi )+\left(2 \zeta _0(\xi )+\frac{\beta '(\xi )}{\beta (\xi )}\right) \zeta_1 (\xi )+\zeta _0'(\xi )+ \zeta_1(\xi)^2=0\,.
\end{align}
This is essentially a transformation of a linear second order equation into a non-linear first order one.

The boundary condition at large $\xi$ can be obtained from the large-$\xi$ expansion of $A_{\tx}(\xi)$ beyond the one of eq. (\ref{axxibdysol}),
\begin{align}
	A_{\tx}(\xi)=\sqrt{\tlrho}e^{-i \bom \sqrt{\tlrho}\frac{1}{4}\left( \ln[2]+\frac{\pi }{2}\right)}\frac{ e^{i \xi  \bar{\omega }} }{\xi}\left(1-\frac{\left(1-\Psi _0^2\right){}^2}{4 \xi ^4}\right)\,,
\end{align}
which gives us 
\begin{align}
	\zeta_0(\xi)+\zeta_1(\xi) = \dfrac{d}{d\xi}\ln(A_{\tx}(\xi)) \approx i \bar{\omega }-\frac{1}{\xi }+\frac{1-2 \Psi _0^2+\Psi _0^4}{\xi ^5}\cdots\,.
\end{align}
By matching this with the large $\xi$ expansion of $\zeta_0(\xi)$,
\begin{align}
	\zeta_0(\xi) \approx \pm i \bar{\omega } -\frac{1}{\xi }-\frac{i}{2 \bar{\omega } \xi ^2}-\frac{i}{8 \bar{\omega }^3 \xi ^4}-\frac{i}{16 \bar{\omega }^5 \xi ^6}\cdots\,,
\end{align}
we know that we should choose the positive root for $\zeta_0(\xi)$ and the resulting large-$\xi$ expansion of $\zeta_1(\xi)$ is given by
\begin{align}
	\zeta_1(\xi) = \frac{i}{2 \bar{\omega } \xi ^2}+\frac{i}{8 \bar{\omega }^3 \xi ^4}\cdots\,,
\end{align}
which is taken as boundary condition for \reef{zetaeom}. 
It is further simplified as we can assume that $\zeta_1(\xi)$ is small compared to $\zeta_0$ and $\beta'/\beta$ such that \reef{zetaeom} reduces to first order in our desired limit with the solution
\begin{equation}
\zeta_1(\xi) \ = \ e^{-  \int_{\Lambda_1}^{\xi} d\tilde{\xi}  \left(2 \zeta _0(\tilde{\xi} )+\frac{\beta '(\tilde{\xi} )}{\beta (\tilde{\xi} )}\right)}\left(C_\zeta + \int_{\Lambda_2}^\xi d\dot{\xi}
e^{ \int_{\Lambda_1}^{\dot{\xi}} d\tilde{\xi}  \left(2 \zeta _0(\tilde{\xi} )+\frac{\beta '(\tilde{\xi} )}{\beta (\tilde{\xi} )}\right)}
 \zeta_0'(\dot{\xi})\right) \ ,
\end{equation}
where the integration constant is fixed by the boundary condition to $C_\zeta = \zeta_1(\Lambda_1) \sim \frac{i}{2 \bar{\omega } \Lambda_1 ^2}+\cdots$.

Given a solution for $\zeta_1$, the solution for $A_{\tx}(\xi)$ is then from the Ansatz \reef{aintansatz}
\begin{align}\label{zetaexpand}
	A_{\tx}(\xi)& = C e^{\int^{\xi}_{\Lambda}\zeta_0(\xi')\,d\xi'} e^{\int^{\xi}_{\Lambda}\zeta_1(\xi')\,d\xi'} \notag\\
	& \approx C e^{\int^{\xi}_{\Lambda}\zeta_0(\xi')\,d\xi'}(1 + \int^{\xi}_{\Lambda} \zeta_1(\xi')\,d\xi')\,,
\end{align}
where $\lambda$ is some large value of $\xi$.
Our approximation holds for $\bom \gg 1$ when $\int^{\xi}_{\Lambda} \zeta_1(\xi')\,d\xi'$ is  small for some finite $\xi$. When $\xi$ approaches 0, the incoming and outgoing modes decouple such that \reef{zetaexpand} is still valid. This can be easily seen in eqs. \reef{axxieom} or \reef{modeeom} as the coupling is proportional to $\partial_\xi (\alpha(\xi)\beta(\xi))$ which actually vanishes in both (large and small) limits of $\xi$.
 Thus, we could find the locations of poles by demanding $\int^{0}_{+\infty} \zeta_1(\xi')\,d\xi' = -1$. 

In practice, however, we only solve \reef{zetaeom} for real $\bom$, then we fit the real part of $\int^{0}_{+\infty}\zeta_1(\xi')\,d\xi'$ into the real part of a generic complex model 
\begin{align}
\Re \left( \int^{0}_{+\infty} \zeta_1(\xi')\,d\xi' \right) \ = \ \left(a\, e^{b\, \bar{\omega }} \bar{\omega }^{c}+ f(\bom)\right)\dfrac{1}{\bom^3}\,,
\end{align}
where $a,b$ and $c$ are complex parameters, $f(\bom)$ is a non-periodic function and asymptotically proportional to $\bom$. 
Then, we can solve this model for $\int^{0}_{+\infty} \zeta_1(\xi')\,d\xi' = -1$ in the complex frequency plane to find 
the locations of poles in the complex frequency plane. One may obtain same result by fitting imaginary part accordingly and the result should also be independent of the particulars of the model. In particular, at large $\bom$, the first term is the most generic one and as $\bom \rightarrow \infty$ we are only interested in the parameter $b$ to give us the tower of excitations.

\begin{figure}[H]
\centering
\includegraphics[width = 0.9\textwidth]{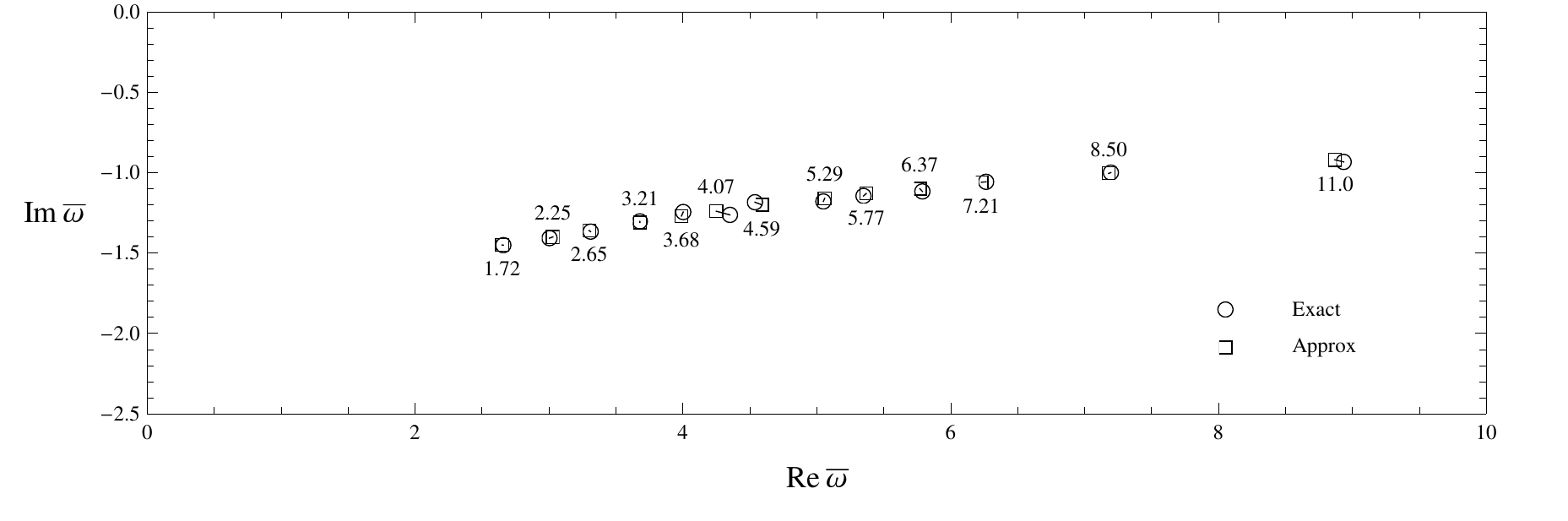} 
\caption{The comparison between the exact locations of few first order poles (circle) and the approximate locations obtained from the analytic continuation method (square). The numbers are the corresponding values of $\barm$.} \label{compole}
\end{figure}

For validation of our procedure, we compare our results for each the first pole at various values of the mass in the range at which we can find the exact location to the exact result in fig. \ref{compole}. Even though the first poles are expected to be the least reliable ones and most dependent on the model, they match with the exact locations very well.

\subsubsection{Results}\label{resres}
Now that we have obtained the locations of the poles of the two-point function -- exactly for large $\barm$ and with a good approximation also for small $\barm$ -- we can use them to analyze the quasiparticle spectrum of our theory in the field theory side.
To do so, we parametrize the poles corresponding to a fixed value of $\bar{m}$ asymptotically (using the highest few excitations) as 
\begin{align}\label{model}
	\bom_n (\bar{m} ) = \bom_0(\barm) n + \bar{\nu}(\bar{m})
\end{align}
for some complex parameters $\bom_0(\barm)$ and $\bar{\nu}(\barm)$ that represent the spacing of excitation levels (and their inverse lifetimes) and some ``ground state energy'' which will obviously not be exactly the value of the $0^{th}$ excited state. The other way round, one can interpret the real part of $1/\bom_0(\barm)$ as some induced length scale that depends on $\bar{m}$, i.e. on the ratio $\sqrt{\rho}/m$.

\begin{figure}[H]
\centering
\includegraphics[width = 0.55\textwidth]{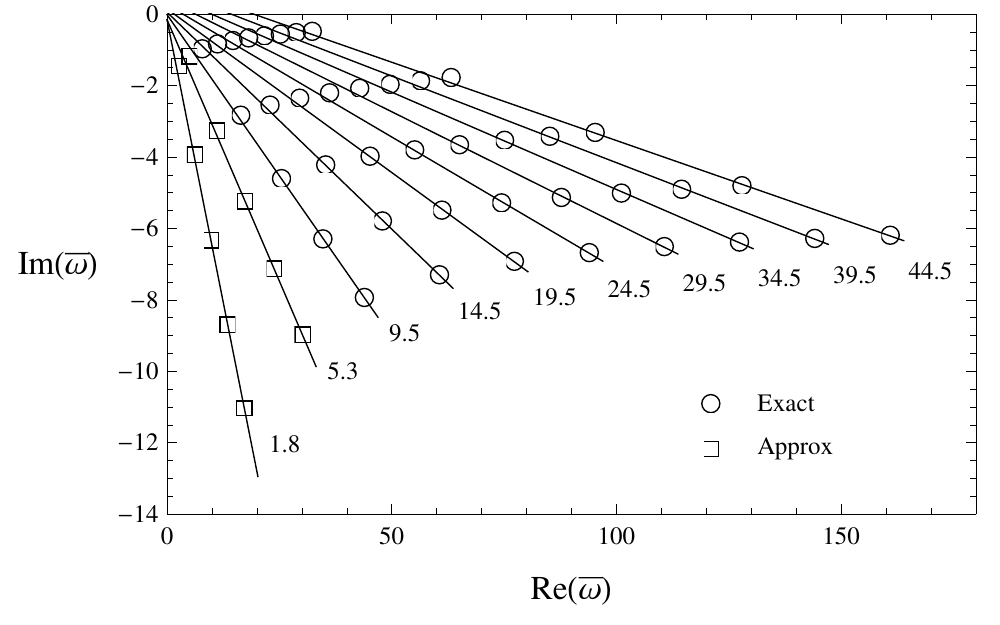}
\caption{The locations of the first five orders of poles obtained from directly solving the E.O.M. for $A_{\tx}(\xi)$ \protect \reef{axxieom} (circles) and from the analytic continuation method (square) compared to our parametrization \protect \reef{model} shown as solid lines. The numbers in the graph are the corresponding values of $\barm$.} \label{poles}
\end{figure}

That this parametrization is well-motivated can be seen in fig. \ref{poles}, where we show the first few excitations for selected values of $\bar{m}$.
We see that parametrization is actually a very good fit, i.e. we have an almost linear tower of excitations, or an induced length scale that persists at high energies. We also find already now that both the excited states and the ground states get heavier and more stable with increasing quark mass. 

\begin{figure}[H]
\centering
\includegraphics[width = 0.45\textwidth]{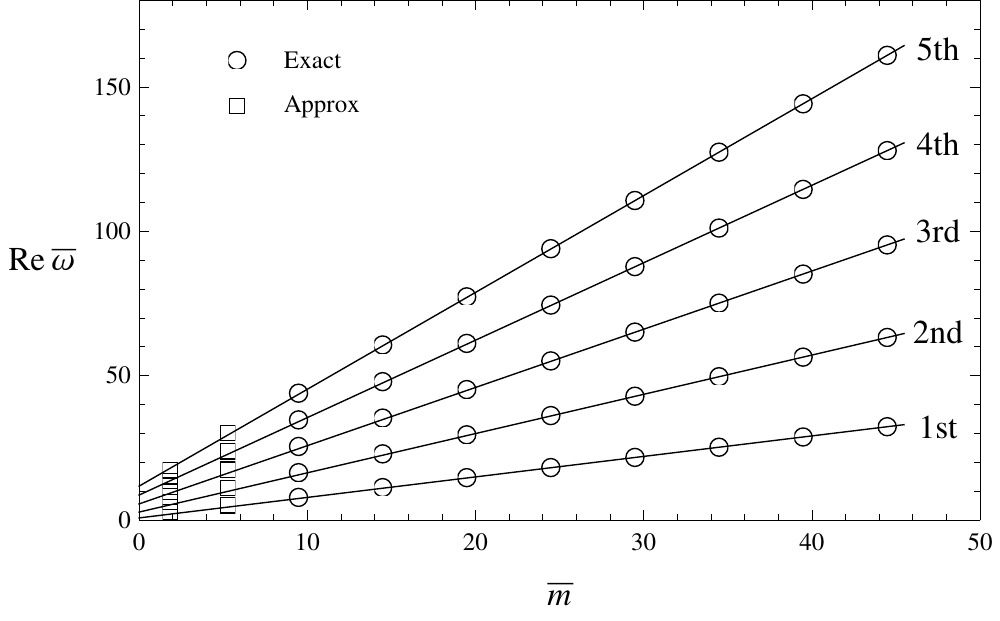} 
\includegraphics[width = 0.45\textwidth]{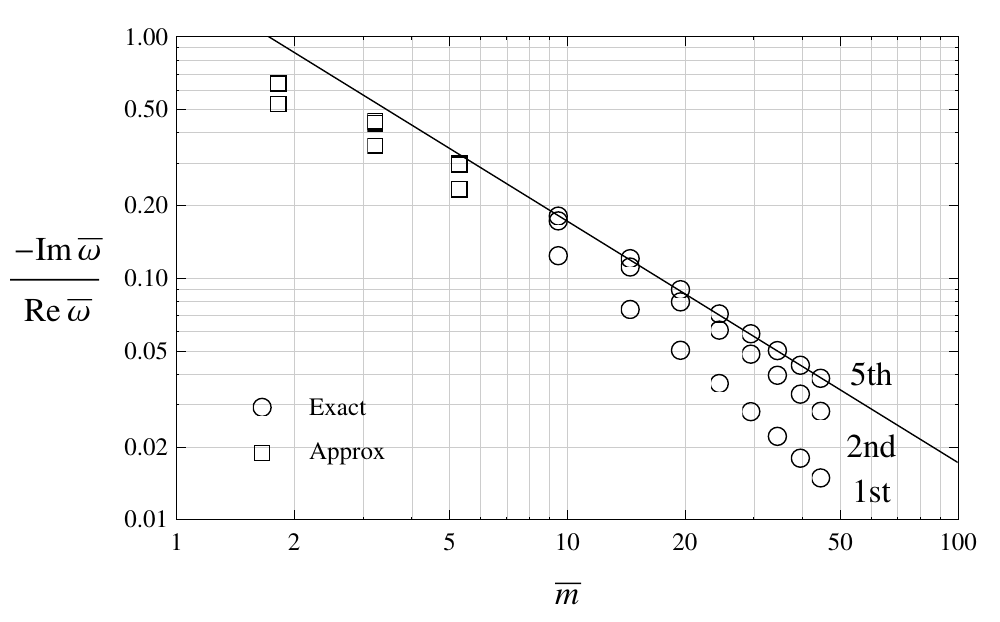}
\caption{Left: The real parts of the locations of poles as functions of $\barm $ with linear fits shown as solid lines. Right: The double-logarithmic plot of $-\Im(\bom_n)/\Re(\bom_n)$ as a function of $\barm$. The solid line indicates the asymptotic behavior $1.72/\barm$ of the 5th order poles at large $\barm$. The numbers indicate the order of the resonances and for the sake of clarity, the 3rd and 4th order poles which lie between the 5th and 2nd are not shown in the graph.} \label{masstowidth}
\end{figure}

To look at the data from a different perspective, we also show the energy of the quasiparticle states, i.e. the real part of $\bom_n$, and their inverse lifetime to mass ratio as a function of the mass in fig. \ref{masstowidth}.
%
%
We see that except for a small deviation at small mass (or large density), the energy of the excitations depends approximately linearly on the the quark mass. The inverse lifetime to mass ratio decreases inversely with increasing mass and, for the higher orders of the excitations, does not depend on the excitation number, such that the asymptotic behavior is approximately $1.72/\barm$.

\begin{figure}[H]
\centering
\includegraphics[width = 0.45\textwidth]{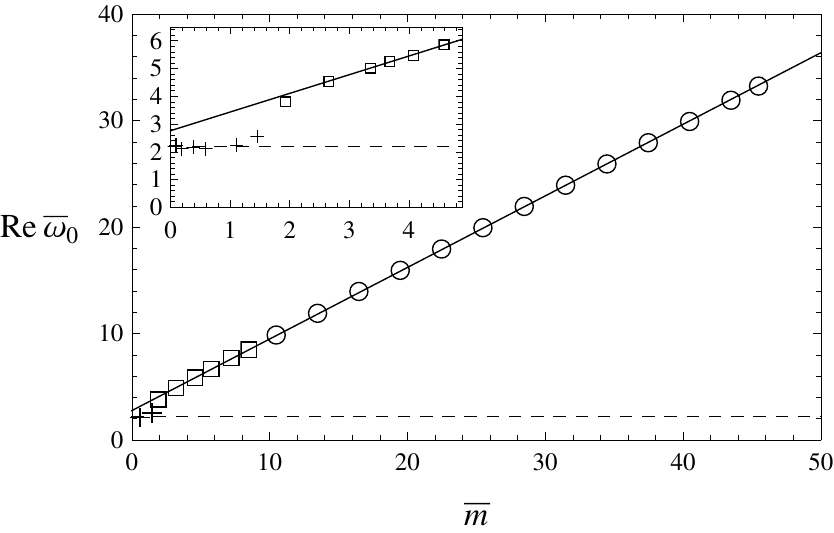}
\hspace{0\textwidth}
\includegraphics[width = 0.49\textwidth]{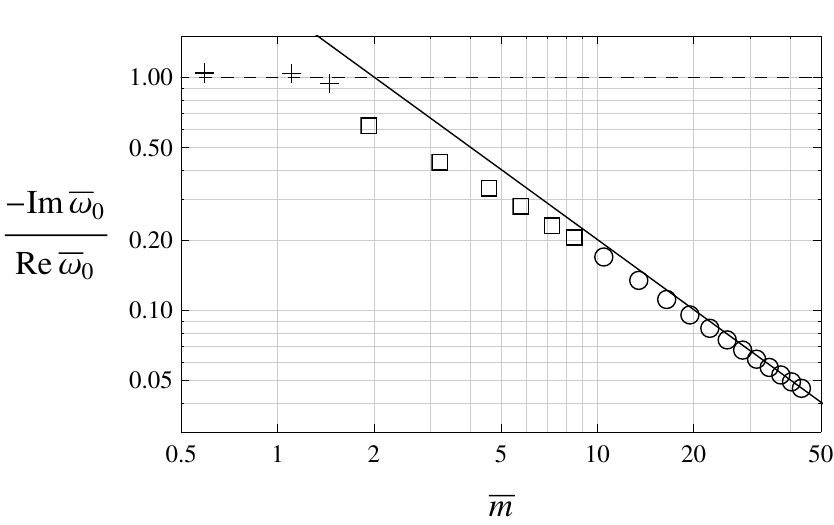} \\
\includegraphics[width = 0.49\textwidth]{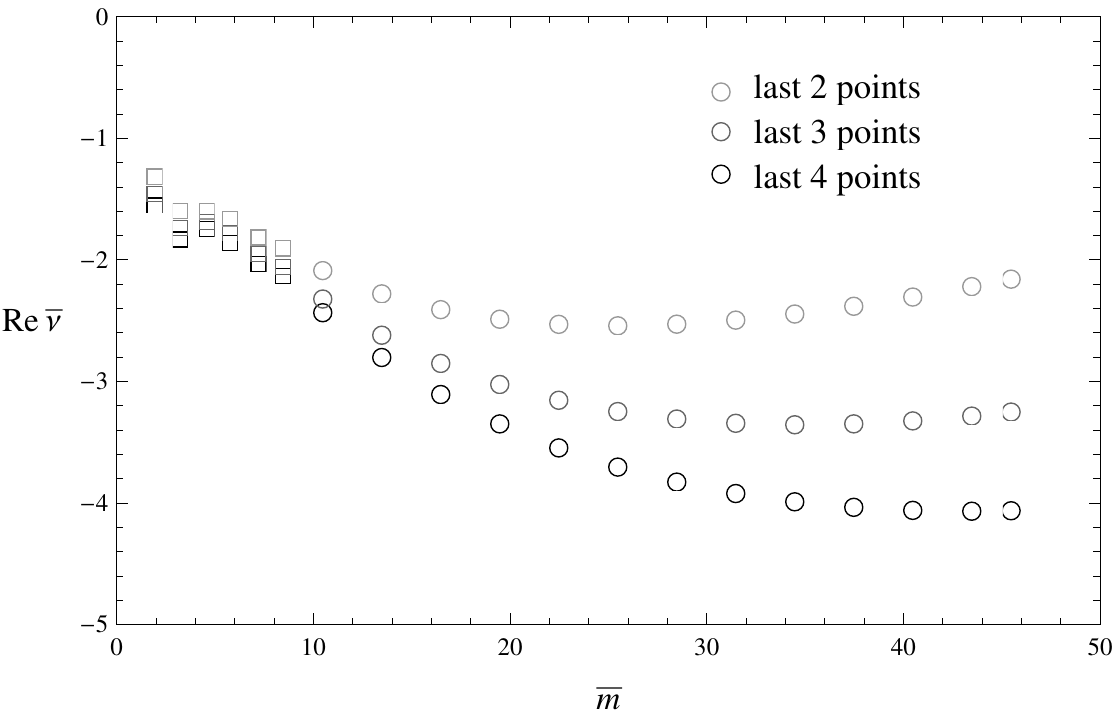}
\hspace{0\textwidth}
\includegraphics[width = 0.49\textwidth]{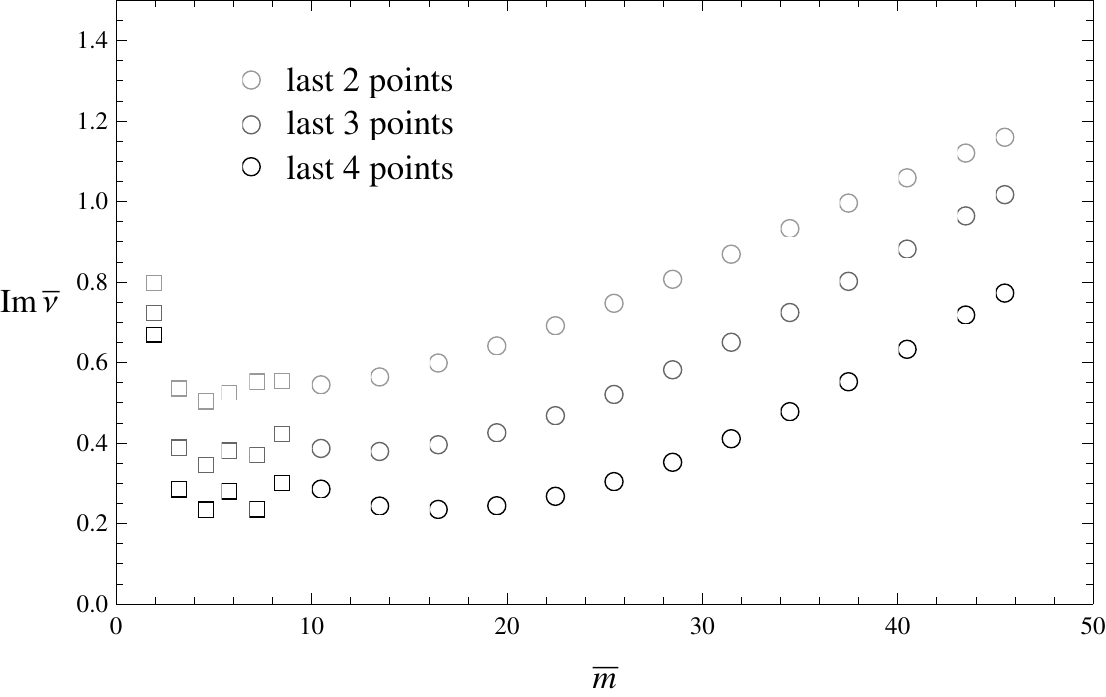} 
\caption{Top left: The ``separation" of quasi-particles as a function of $\barm$, the solid line is the linear behavior obtained at large $\barm$. The crosses correspond to values obtained with the fitting method explaind in the text. The solid and dashed lines show the limiting behavior in the large- and small mass regions. Top right: The ``width to mass" ratio as a function of $\barm$, the solid line indicates the asymptotic behavior $2/\barm$ for large $\barm$. The plots on the bottom show the overall shift in the quasiparticles. To illustrate the level of reliability for the latter, we show the fits for the highest 2, 3 and 4 resonances.}\label{modelpara}
\end{figure}

%
%
%
%
This behavior can be best quantified in the parameters of eq. \ref{model}, that we show in fig. \ref{modelpara}.

To explore the behavior or $\bom_0$ at small masses that are out of reach of the method of sec. \ref{analcon}, we extract the data for $\int^{0}_{+\infty}\zeta_1(\xi')\,d\xi'$ from the conductivity. From eq. \reef{zetaexpand}, we get 
$ \frac{\partial_x A_{\tx}(\xi)}{A_{\tx}(\xi)} = \zeta_0(\xi)+ \zeta_1(\xi)$, and since $\lim_{\xi \rightarrow 0}\zeta_0(\xi) = i \bom$, all the imaginary part of the conductivity originates from $\zeta_1$. As most of the contribution to the integral $\int \zeta_1$ comes from the region at small $\xi$ in which the two modes have decoupled already, we assume that a fit to $\lim_{\xi \rightarrow 0}\zeta_1(\xi)$, i.e. to the imaginary part of the conductivity gives a good indication for the behavior of $\int \zeta_1$ and we perform a fit of the form $\Re A e^{2 \pi i \bom/\bom_0}$ to extract $\bom_0$. Interestingly, this method is reliable at small values of $\bar{m}$ but not at large masses, and it is most reliable at $\bar{m} = 0$ with $\bom_0 \sim 2.20 + 2.20 i$. 
The energy gap between the excitations, $\Re \bom_0$ scales at large masses linearly with the mass as $\omega_0 \sim 12.0 \sqrt{\frac{\rho}{N_f N_c}}+ 4.1 \frac{M_q}{\sqrt{\lambda}}$. On the other hand, the overall shift $\bar{\nu}$ appears approximately constant, so it is not interesting for the rest of this discussion. Also, as we fit asymptotically, for the higher excitations, the values are not reliable, as one can see from the dependence on the number of points take into the fit. 
The linear behavior with the factor of $\sqrt{\lambda}$ suggests that in the limit of large $\bar{m}$, energy levels of the quasiparticles are set by the meson mass scale that was found in \cite{robfirst}, which suggests that the excitations, from fig. \ref{poles} we see even their ground states, either are now mesons, or at least that the there is scattering by mesons.
For small mass, $\bom_0$ approaches within the accuracy of the fittings the more precise value of $\bom_0 \sim 2.20 + 2.20 i$ at $\bar{m} = 0$, such that the actual energy scale scales as $\omega_0 \sim 9.77 \sqrt{\frac{\rho}{N_c N_f}}$, so one of the open questions of \cite{fancycon} is actually resolved. Keeping in mind that $\rho$ is actually the quark number density, this scaling suggests that the quark density arises mostly from baryons that again have plasmon excitations -- density fluctuations that play a role in the context of condensed matter physics (see e.g. \cite{cmtbook}) and are a pure quantum effect.
As we already saw in fig. \ref{masstowidth}, the inverse lifetime to mass ratio starts with a finite value and decreases approximately as $1.95 \frac{2 \pi \sqrt{\lambda}}{M_q}$, where the prefactor depends slightly on how the date is fit and 2 might be a limiting factor. The finiteness of this ratio just means that in all cases the lifetime decreases inversely with the excitation level. While in the plasmonic regime it is obviously independent from the quark mass, in the mesonic regime it increases proportional to $\bar{m}$ which we can now interpret as the ratio of the meson mass to the plasmon mass, or the meson mass to the energy scale related to the geometric distance between the baryons. This also underlines the quantum nature of the decay of the mesonic excitations. We will discuss where this phenomenological appearance of baryons and mesons originates from in the conclusions.

\section{Discussion}\label{conclude}
In this paper, we used the holographic technique to study various properties of fundamental matter with finite mass and baryon number density in the low-temperature regime, coupled to the usual $N_c \gg 1, N = 4$ SYM theory above the deconfinement phase transition. This state of matter might be similar in some aspects for example to what one finds in neutron stars. The fact that we used a top-down approach guarantees us that the field theory is consistent and well defined, giving more relevance to our findings. From the 2-point functions, we obtained on the one hand the AC conductivity over the whole frequency range in sec. \ref{accon}, and on the other hand, we studied the underlying physical parameters -- the relaxation time in the low-frequency Drude limit in sec. \ref{drudesec} and the quasiparticle spectrum and the nature of higher excitations in the large-frequency regime in \ref{quasiparticle}.

Using methods based on the approach in \cite{funnel}, we obtained results that are intrinsically temperature-independent except for the trivial dimensional temperature scaling. By a sensible choice of coordinates and taking the limit of a large quark number density to temperature-squared ratio in sections \ref{setup} and \ref{solution}, we could eliminate the explicit density and temperature dependence from the equations of motion. Hence, the horizon, i.e. the temperature, appears only implicitly as zero-radius boundary condition and the density appears only implicitly through the density to mass-squared ratio. The relevant ratio is $\frac{\rho}{N_c N_f}\frac{\lambda}{M_q^2}$ so the relevant parameters are actually the baryon density and meson mass scale rather than quark number density and quark mass. The role of the temperature as a dimensionful scaling parameter was replaced by the density. 

Looking first at the AC conductivity in sec. \ref{accon}, we found instead of a drude peak a  plateau that is increasingly suppressed with increasing mass, and then a set of resonances appearing with increasing mass before the conductivity reaches the conformal value \cite{conpaper,witten,dolan} at large frequencies. The plateau however appears not to be related to an energy gap due to the lack of an exponential suppression factor. Furthermore, from the Kramers-Kronig relations, we could extract and validate in sec. \ref{drude} the information of the Drude peak, that appears in our limit only as a delta function at zero frequency. We found an exact expression for the relaxation time that scales at small masses or large densities proportional to $\frac{1}{T^2}\sqrt{\frac{\rho}{N_f N_c}}$, so it actually increases with increasing density, which may be due to strong coupling as the fluctuations that carry the current are then more stable at large densities. At large quark mass or small density however the relaxation time scales as $\frac{1}{T^2}\frac{M_q}{\sqrt{\lambda}}$, which is again not a classical mean free path type scaling.

To obtain the underlying quasiparticle spectrum behind the resonances, we employed again a special scaling behavior to obtain a closed-form solution at large frequencies, and analytic continuation  in sec. \ref{analcon} to obtain the particle energies and lifetimes from the quasinormal modes in the gravity side. We saw in sec. \ref{resres} that we have an approximately linear tower of excitations with energy-independent width to mass ratio. At large mass, the energy scale of the excitations was set by the meson mass, whereas at small mass, it was set by $\sqrt{\frac{\rho}{N_c N_f}}$ which could be interpreted as the energy due to the geometric length scale of the baryon number density. The inverse lifetime to mass ratio at large mass was approximately  proportional to $\sqrt{\frac{\rho}{N_c N_f}}\frac{\sqrt{\lambda}}{M_q}$ and independent of the temperature. The latter implies that this lifetime is a pure quantum effect. At small mass however, it approaches a finite number. 

These scalings suggest that the quarks form at large density (small mass) baryons that have collective quantum excitations and at small density (large mass) the system is dominated by mesons. As discussed in \cite{fancytherm,funnel}, we do not have phase transitions, but we are rather in the limits of an equilibrium between (not necessarily stable) baryons, mesons and free quarks. On the gravity side, this means that the embeddings of the probe branes converge towards the probe branes with Minkowski embeddings or probe branes of a baryon vertex (in a confining background), respectively. 
\begin{figure}[h]
\centering
\includegraphics[width = 0.45\textwidth]{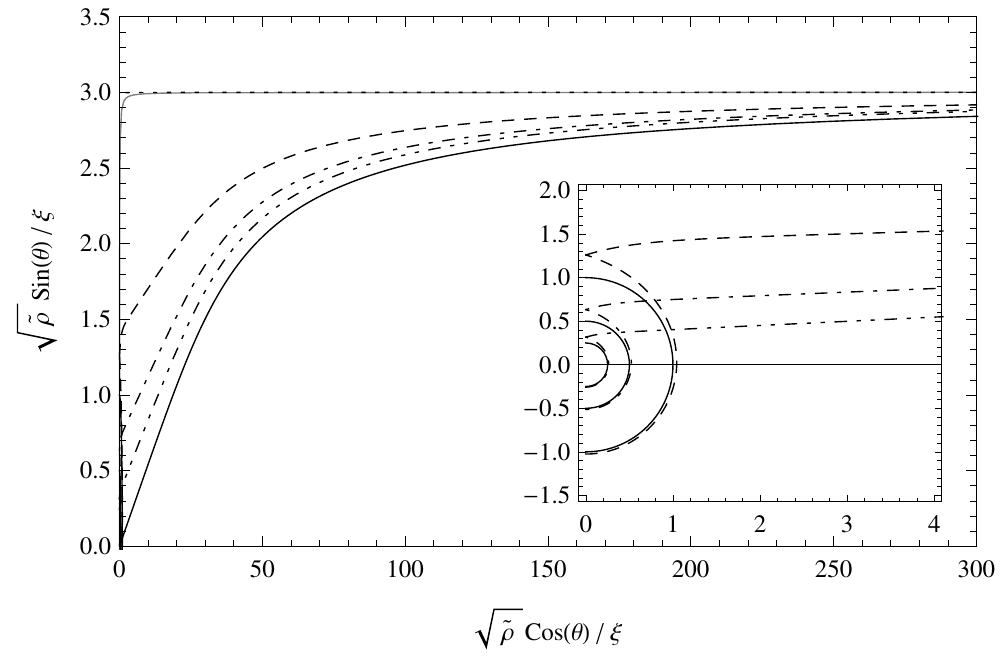} 
\caption{The embeddings of the probe branes at $\tilde m = 3$ and $\bar m = 10$ ($\tilde \rho = 0.09$, gray) and $\bar m = 0.1$ ($\tilde \rho = 900$, black) compared to the Minkowski embedding of the mesonic phase ($\tilde \rho = 0$, dotted) and the baryon vertex embedding corresponding to $\bar m = 0.1$ for a confinig background with $r_{KK}$ of $r_0$ (dashed),  $r_0/2$ (dash-dot) and $r_0/4$ (dash-dot-dot). The inset shows the baryon vertex and the KK-radius.} \label{lastplot}
\end{figure}
We illustrate this in fig. \ref{lastplot}, where we show the embeddings obtained in the small-temperature limit, eq. \ref{psieomxi}, for $\bar m=0.1$ (solid black) and $\bar m=10$ (solid gray), scaled such that the axes correspond to $\frac{r \cos \theta}{r_0} = \frac{\sqrt{\tilde\rho}\cos \theta}{\xi}$  and $\frac{r \sin \theta}{r_0} = \frac{\sqrt{\tilde\rho}\sin \theta}{\xi}$  with a mass of $\tilde m = 3$ and $\tilde \rho = 900$ and $\tilde \rho = 0.09$, respectively. For the small density, we compare it to the Minkowski embedding (black dotted) with $\tilde \rho =0$ and $\tilde m = 3$, that is the dual of the mesonic phase \cite{robfirst,johannaphase}. The similarity in this regime is well-known \cite{long,fancytherm}. Stable baryons correspond to branes wrapping the KK wall in a confined background \cite{wittenbaryon}, forming a baryon vertex to which probe branes are attached and which source the electric field on the probe branes -- in deconfining backgrounds like ours, a baryon vertex does not exist \cite{yunseokbaryon}.  Now, the relevant scale is not the temperature (black hole radius $r_0$) but the KK-scale, set by $r_{KK}$, which represents the deconfinement temperature.  Following the instructions in \cite{callanbaryon} and \cite{yunseokbaryon}, we show the baryon vertex embeddings corresponding to $\tilde m = 3$ and $\tilde \rho = 900$ for confining backgrounds with $r_{KK} = r_0$ (dashed), $r_{KK}= r_0/2$ (dash-dot) and $r_{KK}= r_0/4$ (dash-dot-dot), motivated by the fact that we consider a deconfined background, i.e. we are above the deconfinement phase transition. We see that as $r_{KK}$ decreases, the embedding of the baryon probe brane approaches our embedding in the case of large densities -- supporting the interpretation that the system is dominated by baryons in this limit. It is not quite clear, however, whether between these extreme limits there still free quarks, or whether the system consists entirely of baryons and mesons.
\acknowledgments
The authors would like to thank Yun-Seok Seo for essential help and discussions regarding the baryon vertex and Shesansu Pal and Hyun Cheol Lee for helpful discussions and useful comments. This work was supported by the National Research Foundation of Korea(NRF) grant
funded by the Korea government(MEST) through the Center for Quantum Spacetime(CQUeST)
of Sogang University with grant number 2005-0049409.

\bibliography{bibfile}
\bibliographystyle{JHEP}

\end{document}